\documentclass[10pt,english,a4paper,showpacs,amsmath,amssymb,floatfix]{revtex4-1}
\usepackage{amsmath,amstext,amssymb,graphicx,babel,geometry,setspace,bm,dcolumn,color}
\begin{document}

\title{Emergent Bistability : Effects of Additive and Multiplicative Noise}
\author{Sayantari Ghosh}
\author{Subhasis Banerjee}
\author{Indrani Bose}
\email{indrani@bosemain.boseinst.ac.in}
\affiliation{Department of Physics, Bose Institute, 93/1 Acharya Prafulla Chandra Road, Kolkata - 700009, India}

\begin{abstract}
Positive feedback and cooperativity in the regulation of gene expression
are generally considered to be necessary for obtaining bistable expression
states. Recently, a novel mechanism of bistability termed emergent
bistability has been proposed which involves only positive feedback
and no cooperativity in the regulation. An additional positive feedback
loop is effectively generated due to the inhibition of cellular growth
by the synthesized proteins. The mechanism, demonstrated for a synthetic
circuit, may be prevalent in natural systems also as some recent experimental
results appear to suggest. In this paper, we study the effects of
additive and multiplicative noise on the dynamics governing emergent
bistability. The calculational scheme employed is based on the Langevin
and Fokker-Planck formalisms. The steady state probability distributions
of protein levels and the mean first passage times are computed for
different noise strengths and system parameters. In the region of
bistability, the bimodal probability distribution is shown to be a
linear combination of a lognormal and a Gaussian distribution. The
variances of the individual distributions and the relative weights
of the distributions are further calculated for varying noise strengths
and system parameters. The experimental relevance of the model results
is also pointed out. 
\end{abstract}

\maketitle

\section*{I. INTRODUCTION}

Gene expression, a fundamental activity in the living cell, is a complex
sequence of events resulting in protein synthesis \cite{key-1}. In
the case of deterministic time evolution, the temporal rate of change
of protein concentration is given by the difference between the rates
of synthesis and decay of proteins. When these rates balance each
other the net rate of change is zero and one obtains a steady state
described by a fixed protein concentration. In the case of positively
regulated gene expression, the dynamics may result in bistability,
i.e., the existence of two stable steady states for the same parameter
values \cite{key-2,key-3,key-4}. Bistability, in general, is an outcome
of dynamics involving positive feedback and sufficient nonlinearity.
One way of achieving the latter condition is when multiple bindings
of regulatory molecules occur at the promoter region of the gene (cooperativity
in regulation) or when the regulatory proteins form multimers like
dimers and tetramers which then bind the specific regions of the DNA
\cite{key-2,key-3,key-4,key-5}. The simplest example of bistability
in gene expression is that of a gene the protein product of which
promotes its own synthesis \cite{key-5,key-6}. Positive autoregulation
occurs via the binding of protein dimers at the promoter region resulting
in the activation of gene expression. Experimental evidence of bistability
has been obtained in a wide range of biological systems, e. g. , the
lysis-lysogeny genetic circuit in bacteriophage $\lambda$ \cite{key-7},
the lactose utilization network in \emph{E. coli} \cite{key-8}, the
network of coupled positive feedback loops governing the transition
to the mitotic phase of the eukaryotic cell cycle \cite{key-9}, the
development of competence in the soil bacteria \emph{B. subtilis}
\cite{key-4,key-10} and more recently the activation of the stringent
response in mycobacteria subjected to nutrient depletion \cite{key-11,key-12}.
A number of synthetic circuits have also been constructed which exhibit
bistable gene expression under appropriate conditions \cite{key-13,key-14,key-15}.

Recently, Tan et al. \cite{key-16} have proposed a new mechanism
by which bistability arises, termed emergent bistability, in which
a noncooperative positive feedback circuit combined with circuit induced
growth retardation of the embedding cell give rise to two stable expression
states. The novel type of bistability was demonstrated in a synthetic
gene circuit consisting of a single positive feedback loop in which
the protein product X of a gene promotes its own synthesis in a noncooperative
manner. If the circuit is considered in isolation, one obtains monostability,
i.e., a single stable steady state. In the actual system, the production
of protein X has a retarding effect on the growth of the host cell
so that the circuit function is linked to cellular growth. The protein
decay rate has, in general, two components, the natural degradation
rate and the dilution rate due to cell growth. Since the latter is
reduced on increased protein production, a second positive feedback
loop is effectively generated, as illustrated in Fig. 1. An increased
synthesis of protein X leads to a lower dilution rate, i.e., a greater
accumulation of the protein which in turn promotes a higher amount
of protein production. Tan et al. \cite{key-16} developed a mathematical
model to capture the essential dynamics of the system of two positive
feedback loops and showed that in a region of parameter space bistable
gene expression is possible. The rate equation governing the dynamics
of the system is given by \cite{key-16}

\begin{equation}
\frac{dx}{dt}=\frac{\delta+\alpha x}{1+x}-\frac{\phi x}{1+\gamma x}-x\end{equation}

where the variables and the parameters are nondimensionalized with
$x$ being a measure of the protein amount. The parameter $\delta$
is the rate constant associated with basal gene expression, $\alpha$
represents the effective rate constant for protein synthesis, $\phi$
denotes the maximum dilution rate due to cell growth and $\gamma$
is a parameter denoting the `metabolic burden'. When limited resources
are available to the cell, the synthesis of proteins imposes a metabolic
cost, i.e., the availability of resources for cellular growth is reduced.
The form of the nonlinear decay rate (the second term in Eq. (1))
is arrived at using the Monod model \cite{key-17} which takes into
account the effect of resource or nutrient limitation on the growth
of a cell population. An alternative explanation for the origin of
the nonlinear protein decay rate is based on the fact that the synthesis
of a protein may retard cell growth if it is toxic to the cell \cite{key-18}.

The experimental signature of bistability lies in the coexistence
of two subpopulations with low and high protein levels. In the case
of deterministic dynamics, the cellular choice between two stable
expression states is dictated by the previous history of the system
\cite{key-2,key-3,key-4,key-14}. In this picture, if the cells in
a population are in the same initial state, the steady state should
be the same in each cell. The experimental observation of population
heterogeneity in the form of two distinct subpopulations can be explained
once the stochastic nature of gene expression \cite{key-19,key-20}
is taken into account. Bistable gene expression is characterized by
the existence of two stable steady states separated by an unstable
steady state. A transition from the low to the high expression state,
say, is brought about once the fluctuations associated with the low
expression level cross the threshold set by the protein concentration
in the unstable steady state \cite{key-6,key-10,key-21}. Noise-induced
transitions between the stable expression states give rise to a bimodal
distribution in the protein levels. In a landscape picture, the two
stable expression states correspond to the two minima of an expression
potential and the unstable steady state is associated with the top
of the barrier separating the two valleys \cite{key-22}. A number
of examples is now known which illustrate the operation of stochastic
genetic switches between well-defined expression states \cite{key-7,key-10,key-11,key-12,key-23,key-24,key-25,key-26}.
Stochasticity in gene expression has different possible origins, both
intrinsic and extrinsic. The biochemical events (reactions) involved
in gene expression are probabilistic in nature giving rise to fluctuations
(noise) around mean mRNA and protein levels. The randomness in the
timing of a reaction arises from the fact that the reactants have
to collide with each other and the energy barrier separating the reactants
from the product state has to be crossed for the occurrence of the
reaction. The stochastic time evolution of the system can be studied
using the Master Equation (ME) approach \cite{key-27,key-28}. The
ME is a differential equation describing the temporal rate of change
of the probability that the system is in a specific state at time
t. The state at time t is described in terms of the number of biomolecules
(mRNAs, proteins etc.) present in the system at t. The solution of
the ME gives a knowledge of the probability distribution the lower
two moments of which yield the mean and the variance. One is often
interested in the steady state solution of the ME, i.e., when the
temporal rate of change of the probability is zero. The ME has exact,
analytical solutions only in the cases of simple biochemical kinetics.
A rigorous simulation technique based on the Gillespie algorithm \cite{key-29-1}
provides a numerical solution of the ME. The computational cost in
terms of time and computer memory can, however, become prohibitive
as the complexity of the system studied increases. Two approximate
methods for the study of stochastic processes are based on the Langevin
and Fokker-Planck (FP) equations \cite{key-27,key-28}. These equations
are strictly valid in the case of large numbers of molecules so that
a continuous approximation is justified and the system state is defined
in terms of concentrations of molecules rather than numbers. In fact,
both the equations are obtained from the ME in the large molecular
number limit. In the Langevin equation (LE) additive and multiplicative
stochastic terms are added to the rate equation governing the deterministic
dynamics. The corresponding FP equation is a rate equation for the
probability distribution \cite{key-27,key-28}. Noise in the form
of random fluctuations has a non-trivial effect on the gene expression
dynamics involving positive feedback \cite{key-22,key-29,key-30,key-31}.
In this paper, we investigate the effects of additive and multiplicative
noise on the dynamics, described by Eq. (1), with special focus on
emergent bistability. In Sec. II, we first present the bifurcation
analysis of Eq. (1) and then describe the general forms of the Langevin
and FP equations as well as the expression for the steady state probability
distribution of the protein levels in an ensemble of cells. Sec. III
contains the results of our study when only additive as well as when
both additive and multiplicative types of noise are present. In Sec.
IV, the mean first passage times for escape over the potential barrier
are computed. In Sec. V, we discuss the significance of the results
obtained and also make some concluding remarks.

\begin{figure}
\centering{}\includegraphics[scale=0.3]{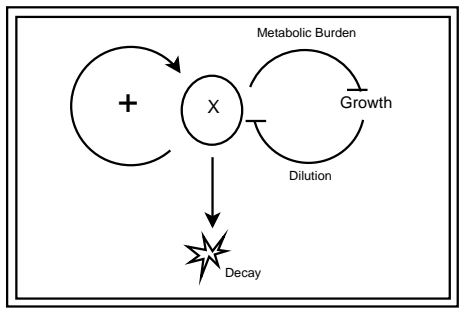}\caption{A genetic circuit illustrating emergent bistability. The protein product
X of a gene promotes its own synthesis as well as retards the growth
of the host cell. Growth retardation combined with a lower dilution
rate of the protein concentration effectively generate a second positive
feedback loop. The arrowhead in a regulatory link denotes activation
and the hammerhead symbol denotes repression. }

\end{figure}

\begin{figure}
\begin{centering}
\includegraphics[scale=0.8]{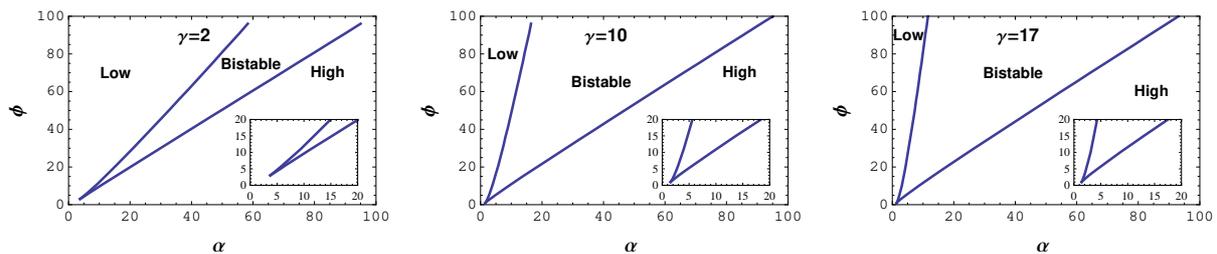}
\par\end{centering}

\centering{}\caption{Bifurcation diagrams corresponding to Eq. (1) in the $\phi$ - $\alpha$
plane with $\gamma=2,10$ and 17 in the successive plots. (inset)
Portions of bifurcation diagrams amplified to indicate how a transition
from the low to the high expression state can be brought about without
passing through the region of bistability. The rate constant $\delta=0.01$. }

\end{figure}

\section*{II. LANGEVIN AND FOKKER-PLANCK EQUATIONS}

In Eq. (1), the steady state condition $\frac{dx}{dt}=0$ results
in bistability, i.e., two stable gene expression states in specific
parameter regions. Fig. 2 represents the bifurcation diagrams in the
$\phi-\alpha$ plane for different values of the metabolic cost parameter
$\gamma$. The region of bistability is bounded by lines at which
bifurcation from bistability to monostability occurs. The steady states
in the region of bistability are the three physical solutions ($x$
real and positive) of the cubic polynomial equation obtained from
Eq. (1) by putting $\frac{dx}{dt}=0$. Two of the solutions correspond
to stable steady states and these are separated by an unstable steady
state represented by the third solution. At the bifurcation point
separating bistability from the monostable low expression state, the
higher stable state solution merges with the unstable steady state
solution. Similarly, at the bifurcation point separating bistability
from the monostable high expression state, the lower two solutions
merge so that away from this point and towards the right in the $\phi$
- $\alpha$ plane only the stable high expression state survives.
The bifurcation analysis has been carried out using the software package
Mathematica. One notes that the extent of the region of bistability
increases as the value of the parameter $\gamma$ increases. When
$\gamma=0$, i.e., the net decay rate of the proteins has the form
$-(\phi+1)x$, there is no region of bistability in the parameter
space. For $\gamma\neq0$, one can pass from a monostable low expression
state across a region of bistability to a monostable high expression
state by increasing $\alpha$ and decreasing $\phi$. One also observes
that one can directly go
from the low to the high expression state without passing through
the region of bistability. The region of parameter space through 
which the bypass can occur is dependent on $\gamma$, diminishing 
in size with increasing values of $\gamma$. We next include appropriate noise terms
in Eq. (1) to investigate the effects of noise on the deterministic
dynamics. 

\begin{figure}
\centering{}\includegraphics[bb=119bp 212bp 492bp 592bp,scale=0.7]{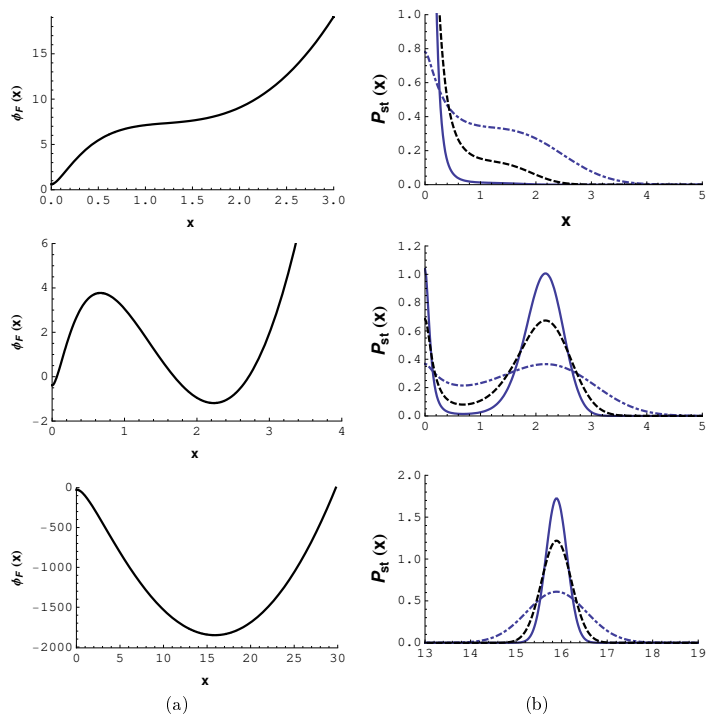}\caption{\textbf{(a) }
Stochastic potential $\phi_{F}(x)$ (Eq. (16)) in the
case of purely additive noise as the parameter $\alpha$ is tuned
from low to high values ($\alpha=5.5,5.96$ and $19$). The other
parameter values are $\gamma=10$, $\delta=0.01$, and $\phi=20$
with the noise strength kept fixed at $D_{2}=0.05$. \textbf{(b) }Steady
state probability distribution, $P_{st}(x)$, for successive values
of $\alpha=5.5,5.96$ and $19$. For each values of $\alpha$, $P_{st}(x)$
is computed and displayed for three different noise strengths $D_{2}=0.05$
(solid line), $D_{2}=0.1$ (dashed line), $D_{2}=0.4$ (dot-dashed
line). }

\end{figure}

The general stochastic formalism based on the Langevin and FP equations
for the steady state analysis of a bistable system is described in
Refs. \cite{key-27,key-28,key-32,key-33,key-34,key-35}. We apply
the formalism to investigate the effects of additive and multiplicative
noise on emergent bistability, the governing equation of which is
described in Eq. (1). We consider an one-variable LE containing a
multiplicative and an additive noise term:\begin{equation}
\frac{dx}{dt}=f(x)+g_{1}(x)\epsilon(t)+\Gamma(t)\end{equation}

where $\epsilon(t)$ and $\Gamma(t)$ represent Gaussian white noises
with mean zero and correlations given by

\begin{eqnarray}
<\epsilon(t)\epsilon(t')> & = & 2D_{1}\delta(t-t') \nonumber \\
<\Gamma(t)\Gamma(t')> & = & 2D_{2}\delta(t-t') \nonumber \\
<\epsilon(t)\Gamma(t')> & = & <\Gamma(t)\epsilon(t')>=2\lambda\sqrt{D_{1}D_{2}}\delta(t-t')
\end{eqnarray}

$D_{1}$ and $D_{2}$ are the strengths of the two types of noise
$\epsilon(t)$ and $\Gamma(t)$ respectively and $\lambda$ is the
degree of correlation between them. The first term, $f(x)$, in Eq.
(2) represents the deterministic dynamics. With the dynamics governed
by Eq. (1), the function $f(x)$ is given by, 

\begin{equation}
f(x)=\frac{\delta+\alpha x}{1+x}-\frac{\phi x}{1+\gamma x}-x\end{equation}

The additive noise $\Gamma(t)$ represents noise arising from an external
perturbative influence or originating from some missing information
embodied in the rate equation approximation \cite{key-22}. Gene expression
consists of the major steps of transcription and translation which
are a complex sequence of biochemical events. Regulation of gene expression
as well as processes like cell growth have considerable influence
on the gene expression dynamics. In many of the models of gene expression,
some of the elementary processes (say, transcription and translation
as in the case of the model described by Eq. (1)) are lumped together
and an effective rate constant associated with the combined process
(e.g., $\alpha$, the effective rate constant for protein synthesis
in Eq. (1)). It is, however, expected that the rate constants fluctuate
in time due to a variety of stochastic influences like fluctuations
in the number of regulatory molecules and RNA polymerases. In the
LE, the fluctuations in the rate constants are taken into account
through the inclusion of multiplicative noise terms like $g_{1}(x)\epsilon(t)$
in Eq. (2). In the present study, we consider two types of multiplicative
noise associated with the rate constants $\alpha$ (effective rate
constant for protein synthesis) and $\phi$ (the maximum dilution
rate). The rates vary stochastically. i.e., $\alpha\rightarrow\alpha+\epsilon(t)$
with

\begin{equation}
g_{1}(x)=\frac{x}{1+x}\end{equation}

in one case and $\phi\rightarrow\phi+\epsilon(t)$ with

\begin{equation}
g_{1}(x)=-\frac{x}{1+\gamma x}\end{equation}

in the other case (see Eq. (1)). As mentioned before, $\epsilon(t)$
represents random fluctuations of the Gaussian white noise-type. There
are alternative versions of the LE in which the noise terms added
to the deterministic part have an explicit structure in terms of gene
expression parameters \cite{key-7,key-29-1,key-32-1}. Gillespie \cite{key-29-1}
has shown how to derive the Chemical LE starting with the Master equation
describing the stochastic time evolution of a set of elementary reactions
with the noise terms depending on the number of molecules as well
as the gene expression parameters. Similarly, for an effective kinetic
equation of the form 

\begin{equation}
\frac{dn}{dt}=f(n)-g(n)\end{equation}

where $n$ is the number of molecules and $f(n)$, $g(n)$ the synthesis
and decay rates respectively, one can write the LE as \cite{key-7,key-32-1}

\begin{equation}
\frac{dn}{dt}=f(n)-g(n)+\sqrt{f(n)+g(n)}\,\Gamma(t)\end{equation}

While these alternative approaches have their own merit, the majority
of stochastic gene expression studies, based on the Langevin formalism,
start with equations of the type shown in (2). The choice is dictated
by the simplicity of the calculational scheme with specific focus
on the separate effects of additive and multiplicative noise (fluctuating
rate constants) on the gene expression dynamics.

The FP equation can be developed from the LE following usual procedure
\cite{key-22,key-28,key-31}. The FP equation corresponding to Eq.
(2) is \cite{key-22,key-28,key-31}

\begin{equation}
\frac{\partial P(x,t)}{\partial t}=-\frac{\partial}{\partial x}[A(x)P(x,t)]+\frac{\partial^{2}}{\partial x^{2}}[B(x)P(x,t)]\end{equation}

where

\begin{equation}
A(x)=f(x)+D_{1}g_{1}(x)g_{1}'(x)+\lambda\sqrt{D_{1}D_{2}}g_{1}'(x)\end{equation}

and

\begin{equation}
B(x)=D_{1}[g_{1}(x)]^{2}+2\lambda\sqrt{D_{1}D_{2}}g_{1}(x)+D_{2}\end{equation}

The steady state probability distribution (SSPD), from Eq. (4), is
given by\cite{key-28,key-32,key-33}

\[
P_{st}(x)=\frac{N}{B(x)}\exp[\int^{x}\frac{A(x)}{B(x)}dx]\]

\begin{equation}
=\frac{N}{\{D_{1}[g_{1}(x)]^{2}+2\lambda\sqrt{D_{1}D_{2}}g_{1}(x)+D_{2}\}^{\frac{1}{2}}}\exp[\int^{x}\frac{f(x')dx'}{D_{1}[g_{1}(x')]^{2}+2\lambda\sqrt{D_{1}D_{2}}g_{1}(x')+D_{2}}]\end{equation}

where N is the normalization constant. Eq. (12) can be recast in the
form

\begin{equation}
P_{st}(x)=Ne^{-\phi_{F}(x)}\end{equation}

with

\begin{equation}
\phi_{F}(x)=\frac{1}{2}\ln[D_{1}[g_{1}(x)]^{2}+2\lambda\sqrt{D_{1}D_{2}}g_{1}(x)+D_{2}]-\int^{x}\frac{f(y)dy}{D_{1}[g_{1}(y)]^{2}+2\lambda\sqrt{D_{1}D_{2}}g_{1}(y)+D_{2}}\end{equation}

$\phi_{F}(x)$ defines the `stochastic potential' corresponding to
the FP equation.

\section*{III. RESULTS FOR ADDITIVE AND MULTIPLICATIVE NOISE}

We first consider the case when only the additive noise term is present
in Eq. (2), i.e., the second term on the  r.h.s. is missing. The function
$f(x)$ is given in Eq. (4). As pointed out in \cite{key-16}, the
parameters $\alpha$ (effective rate constant for protein synthesis)
and $\phi$ (maximum dilution rate) are experimentally tunable. From
Eqs. (12) and (14), the expression potentials, $\phi_{F}(x)$, and
the associated steady state probability distributions, $P_{st}(x)$,
can be computed. The stochastic potential $\phi_{F}(x)$ has the form

\begin{equation}
\phi_{F}(x)=\frac{1}{2}\ln D_{2}-\frac{1}{D_{2}}\int^{x}f(y)dy\end{equation}

\begin{equation}
=\frac{1}{2}\ln D_{2}+\frac{1}{D_{2}}\phi_{D}(x)\end{equation}

where $\phi_{D}(x)$ is the deterministic potential, i.e., $f(x)=-\frac{\partial\phi_{D}(x)}{\partial x}$.
In the presence of only additive noise, the stochastic and deterministic
potentials have similar forms. Figure 3 displays $\phi_{F}(x)$ and
$P_{st}$(x) versus $x$ as the parameter $\alpha$ is tuned from
low to high values ($\alpha=5.5,5.96,19$). The other parameter
values are kept fixed with $\gamma=$10, $\delta=$0.01 and $\phi=$20,
the values being identical to those reported in \cite{key-16}. For
each value of $\alpha$, the potential $\phi_{F}(x)$ is plotted only
for the noise strength $D_{2}$= 0.05 whereas $P_{st}(x)$ is plotted
for the noise strengths $D_{2}$=0.05 (solid line), 0.1 (dashed line)
and 0.4 (dot-dashed line). One finds that by tuning $\alpha$ from
low to high values, one can pass from a monostable low expression
state through a region of bistability, i.e., a coexistence of low
and high expression states to a monostable high expression state.
The region of bistability is distinguished by the appearance of two
prominent peaks in the distribution of protein levels. One can also
keep $\alpha$ fixed and obtain a similar set of plots by varying
the maximum dilution rate $\phi$. 

\begin{figure}
\begin{centering}
\includegraphics[scale=0.8]{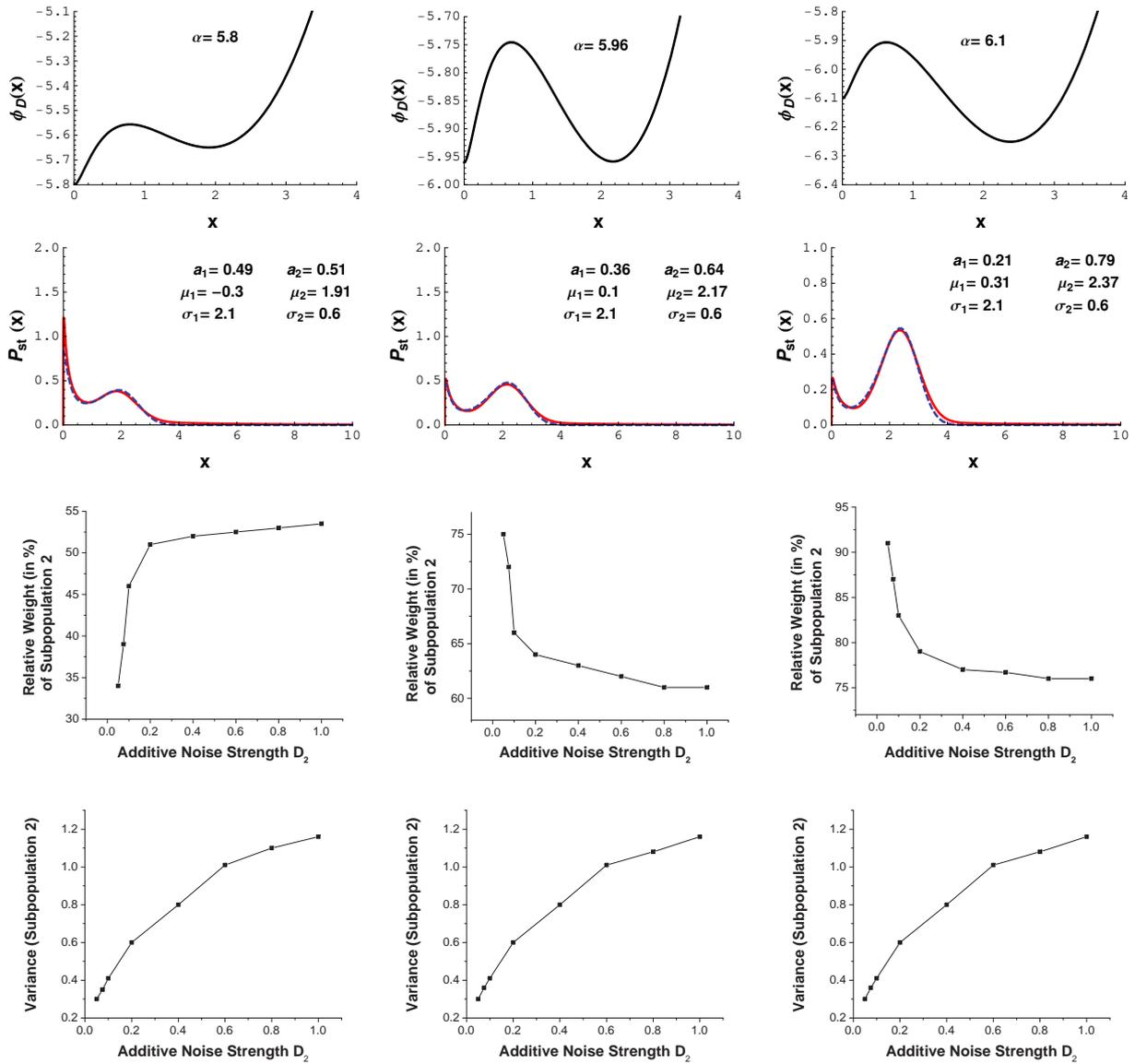}
\par\end{centering}

\caption{Relative weight of Subpopulation 2 and variance of the associated
probability distribution as a function of the additive noise strength
$D_{2}$, computed using a mixture model. (First Panel) Deterministic
potentials for $\alpha=5.8$, $5.96$ and $6.1$ in the region of
bistability. The other parameter values are the same as in the case
of Fig. 3. (Second Panel) Steady state probability distribution $P_{st}(x)$
versus $x$ (solid line) for $\alpha=5.8$, $5.96$ and $6.1$ as
fitted by the distribution (dashed line) obtained from a mixture model.
The individual probability distributions are lognormal (Subpopulation
1) and Gaussian (Subpopulation 2). The parameters of the distributions
and the coefficients $a_{1}$ and $a_{2}$ are also mentioned ($a_{1} + a_{2}=1$). (Third
Panel) Relative weight (in percentage) of Subpopulation 2 versus $D_{2}$.
(Fourth Panel) Variance of Gaussian probability distribution (Subpopulation
2) as a function of $D_{2}$.}

\end{figure}

The stochastic potential $\phi_{F}(x)$ is indicative of the steady
state stability. In Fig. 3(a), for $\alpha=5.5$, the potential has
a deep (shallow) minimum at low (high) values of x. For low values
of the noise strength, e. g. , $D_{2}$=0.05 (solid line), the steady
state probability distribution $P_{st}(x)$ has a single prominent
peak. Noise can induce transitions from one local minimum to the other
of the stochastic potential. The minima represent steady states which
are separated by an energy barrier. For $\alpha=5.5$ and $\alpha=5.96$,
there are two energy barriers corresponding to the transitions from
the low to the high expression states and vice versa. In the case
of $\alpha=5.5$, increased magnitudes of the additive noise flip
the switch in the unfavorable direction (energy barrier higher) so
that $P_{st}(x)$ has a second peak, albeit not prominent, at a higher
expression level. When $\alpha=5.96$, the energy barriers are of
similar magnitude and $P_{st}(x)$ has two prominent peaks at low
and high expression levels when the noise strength is low ($D_{2}=0.05$).
With increased noise strengths, the two expression levels are more
readily destabilized resulting in a smearing of the expression levels.
$P_{st}(x)$ has now a finite value for intermediate values of $x$.
In the case of $\alpha=19$, the stochastic potential has a single
minimum at the high expression level so that $P_{st}(x)$ is unimodal.
With higher values of the noise strength $D_{2}$, the probability
distribution becomes broader. One also notes that with increased values
of $\alpha$, the magnitude of the high expression level also increases. 

\begin{figure}
\begin{centering}
\includegraphics[scale=0.8]{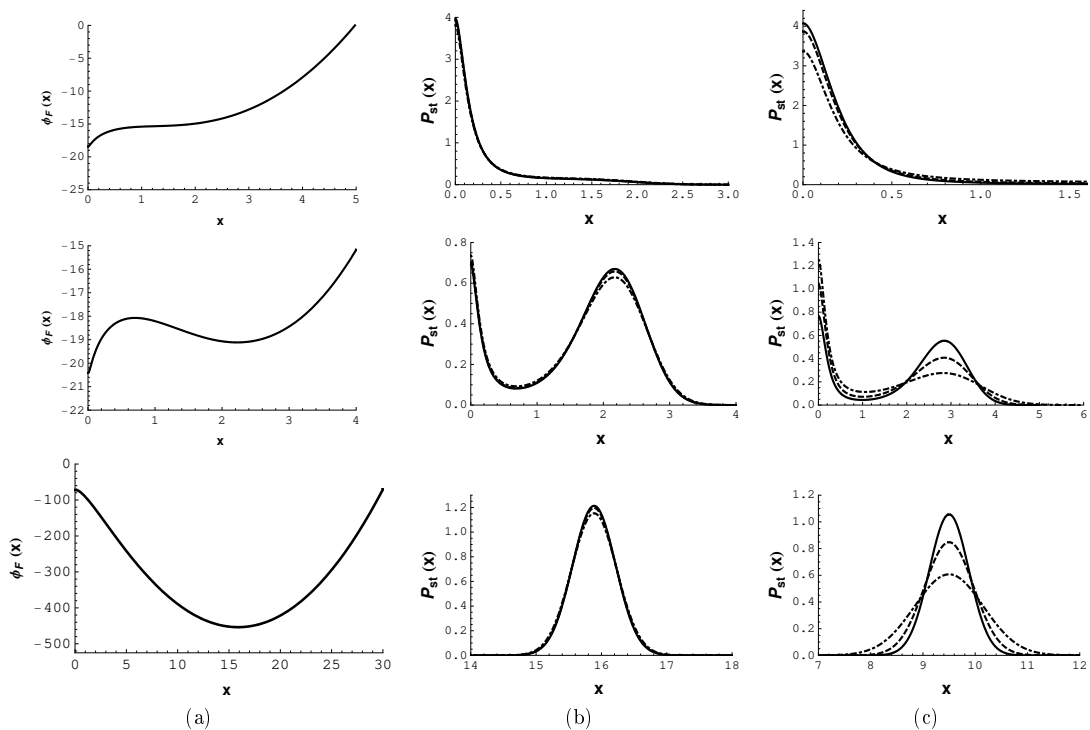}
\par\end{centering}

\centering{}\caption{Both additive and multiplicative noise terms are present with the
multiplicative noise associated with the nonlinear protein decay rate.
Stochastic potential $\phi_{F}(x)$ (a) and the steady state probability
distributions $P_{st}(x)$ ((b) and (c)) are displayed for $\alpha=5.5,5.96$
and $19$ (b) and $\alpha=14.8,15.34$ and $21$(c). The other parameter
values are fixed at $\delta=0.01$, and $\phi=20$ with $\gamma=10$
(b) and $\gamma=2$ (c). The noise strength $D_{2}$ is kept fixed
at $D_{2}=0.1$ whereas the noise strength $D_{1}$ has the values
$D_{1}=0.05$ (solid line), $D_{1}=0.4$ (dashed line), $D_{1}=1.2$
(dot-dashed line) in (b) and (c). The stochastic potentials in (a) are
plotted for $D_{1}=0.05$ and $D_{2}=0.05$ with the other parameter
values as in (b).}

\end{figure}

In the region of bistability, the total population is a mixture of
two subpopulations. In order to determine the effects of additive
noise on the relative weights and variances of the subpopulation probability
distributions, we take recourse to a mixture model in which the steady
state probability distribution for the total population, $P_{st}(x)=a_{1}P_{st1}(x)+a_{2}P_{st2}(x)$.
$P_{st1}(x)$ and $P_{st2}(x)$ are the steady state probability distributions
for Subpopulation 1 and 2 respectively and $a_{1}$, $a_{2}$ are
the coefficients in the linear combination ($a_{1} + a_{2}=1$). Subpopulation 1 (2) is
characterized by predominantly low (high) expression levels. The upper
panel of Fig. 4 shows the deterministic potentials, $\phi_{D}(x)$,
for $\alpha=5.8$, $5.96$ and $6.1$ in the region of bistability.
The other parameter values are the same as in the case of Fig. 3.
The second panel in Fig. 4 shows the steady state probability distributions,
$P_{st}(x)$ (solid line), in the three cases as well as the fitting
distributions (dashed line) using the mixture model. The additive
noise strength $D_{2}$ is kept fixed at the value $D_{2}=0.2$. In
each case, $P_{st1}(x)$ and $P_{st2}(x)$ are given by a lognormal
and a Gaussian distribution respectively with the forms

\begin{center}
$P_{st1}(x)=\frac{e^{-\frac{(log[x]-\mu_{1})^{2}}{2\sigma_{1}^{2}}}}{\sqrt{2\pi}x\sigma_{1}}$ 
\par\end{center}

\begin{flushleft}
with mean $=e^{(\mu_{1}+\frac{\sigma_{1}^{2}}{2})}$ and variance
$=e^{(2\mu_{1}+\sigma_{1}^{2})}\,(e^{\sigma_{1}^{2}}-1)$ and
\par\end{flushleft}

\begin{center}
$P_{st2}(x)=\frac{e^{-\frac{(x-\mu_{2})^{2}}{2\sigma_{2}^{2}}}}{\sqrt{2\pi}\sigma_{2}}$ 
\par\end{center}

\begin{flushleft}
with mean $=\mu_{2}$ and variance $=\mbox{\ensuremath{\sigma}}_{2}^{2}$.
\par\end{flushleft}

The values of the parameters of the individual distributions as well
as the coefficients $a_{1}$ and $a_{2}$ are listed in Fig. 4. The
third panel in the figure shows the relative weight (in percentage)
of the Subpopulation 2 as the additive noise $D_{2}$ is varied for
the three different cases $\alpha=5.8$, $5.96$ and $6.1$. The lowest
panel of Fig. 4 shows the variances of the probability distributions
associated with the Subpopulation 2 versus $D_{2}$ in the three cases.
The plots for Subpopulation 1 are similar in nature. Increased additive
noise strength enhances noise-induced transitions over the potential
barrier. For $\alpha=5.8$, the transitions are from the low to the
high expression state so that the relative weight of Subpopulation
2 increases with increased noise strength. In the other two cases,
the relative weight of Subpopulation 2 decreases with the increase
in the magnitude of $D_{2}$. Increased additive noise strength further
increases the spreading (measured by variance) of the protein distributions
around the mean levels, i.e., brings in greater heterogeneity in the
cell population. The lowest panel of Fig. 4 shows that the variance
is not affected by $\alpha$ but depends only on the additive noise
strength $D_{2}$.

\begin{figure}
\begin{centering}
\includegraphics[bb=120bp 204bp 492bp 588bp,clip,scale=0.7]{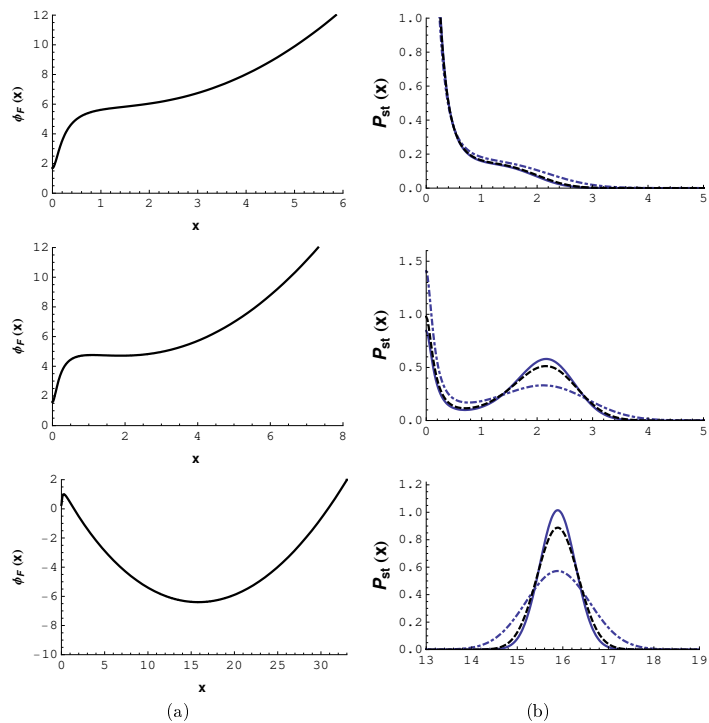}
\par\end{centering}

\centering{}\caption{Both additive and multiplicative noise terms are present with the
multiplicative noise associated with the effective rate constant for
protein synthesis $\alpha$. Stochastic potential $\phi_{F}(x)$ (a)
and the steady state probability distribution $P_{st}(x)$ (b) are
displayed for $\alpha=5.5,5.96$ and $19$. The other parameter values
are fixed at $\gamma=10$, $\delta=0.01$, and $\phi=20$. The noise
strength $D_{2}$ is kept fixed at $D_{2}=0.1$ whereas the noise
strength $D_{1}$ has values $D_{1}=0.05$ (solid line), $D_{1}=0.1$
(dashed line), $D_{1}=0.4$ (dot-dashed line) in (b). }

\end{figure}
 We next consider the case when both additive and multiplicative noise
terms are present in the LE (Eq. (2)). We first assume that the multiplicative
noise is associated with the maximum dilution rate $(\phi\rightarrow\phi+\epsilon(t))$,
i.e., $g_{1}(x)$ in Eq. (2) is given by Eq. (6). We designate this
type of noise as Type 1 multiplicative noise. Also, the two types
of noise are taken to be uncorrelated, i.e., the parameter $\lambda$
(Eq. (3)) is zero. Figure 5 shows the expression potential $\phi_{F}(x)$
(a) and the steady state probability distributions, $P_{st}$(x) ((b)
and (c)) versus $x$ for $\alpha=$ $5.5,5.96$ and 19 (b) and $\alpha=14.8,15.34$
and $21$ (c). The other parameter values are fixed at $\delta=$0.01
and $\phi=$20 with $\gamma=10$ (b) and $\gamma=2$ (c). In computing
$P_{st}(x)$, the additive noise strength $D_{2}$ is kept fixed at
$D_{2}=0.1$ whereas the multiplicative noise strength $D_{1}$ has
the values $D_{1}=0.05$ (solid line), 0.4 (dashed line) and $1.2$
(dot-dashed line). The expression potentials in (a) are plotted for
$D_{1}=0.05$ and $D_{2}=0.05$ with the other parameter values as
in (b). The form of the stochastic potential is given by Eq. (14),
which differs substantially from that of the deterministic potential.
As in the case of Fig. 2, there is a progression from the monostable
low expression state through a region of bistability to the monostable
high expression state. From Fig. 5(b) one finds that the probability
distributions $P_{st}(x)$ versus $x$ are almost unaffected by changing
the multiplicative noise strength $D_{1}$ from $0.05$ to $1.2$.
This is because $g_{1}(x)$, as given in Eq. (6), has a small value
due to the high value of $\gamma$ ($\gamma=10$ ) in the denominator.
Noticeable differences in the probability distributions appear (Fig.
5(c)) when $\gamma$ has a lower value ($\gamma=2$). One can conclude
that increased fluctuations in the maximum dilution rate parameter
$\phi$ have little effect on $P_{st}(x)$ for moderately high values
of $\gamma$. This is certainly not the case when $\gamma=0$ or has
lower values. 

\begin{figure}
\begin{centering}

\par\end{centering}

\begin{centering}
\includegraphics[scale=0.8]{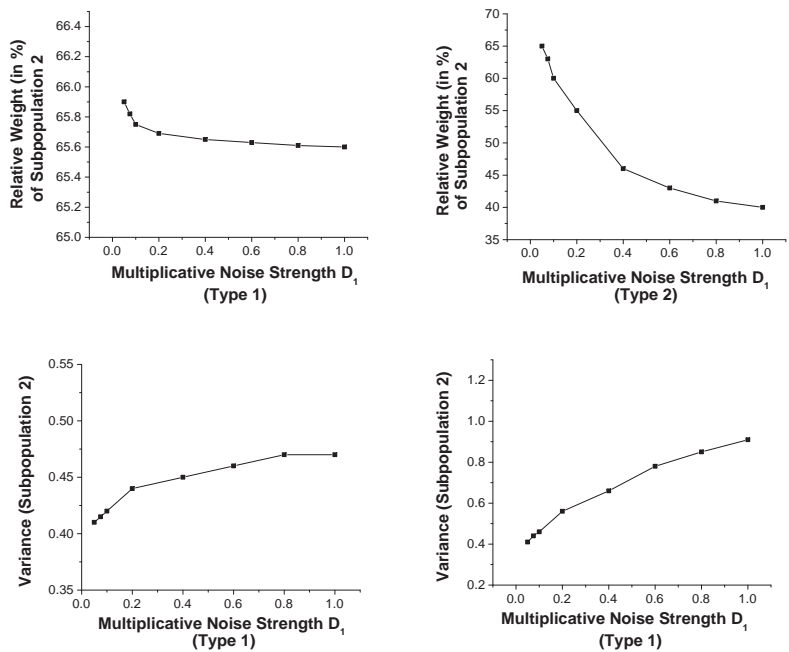}
\par\end{centering}

\caption{Relative weight of Subpopulation 2 and variance of the associated
probability distribution as a function of Type 1 and Type 2 multiplicative
noise strength $D_{1}$, as computed using a mixture model. The additive
noise strength is kept fixed at $D_{2}=0.1$. The value of $\alpha=5.96$
with the other parameter values the same as in the case of Fig. 4.}
 
\end{figure}

Figure 6 exhibits plots similar to Figs. 3 and 5 with the multiplicative
noise term associated with the protein synthesis term, i.e., $g_{1}(x)$
is as given in Eq. (5). We designate this type of noise as Type 2
multiplicative noise. Increased multiplicative noise strengths now
have a greater effect on the steady state probability distributions
than in the earlier case when the multiplicative noise is associated
with the maximum dilution rate $\phi$ (Fig. 5(b)). We again use the
mixture model to determine the relative weights (in percentage) of
the subpopulations as well as their variances as a function of the
multiplicative noise strength $D_{1}$ with the noise being either
of Type 1 (fluctuations in the maximum dilution rate $\phi$) or Type
2 (fluctuations in the effective protein synthesis rate constant $\alpha$).
In both the cases, the additive noise strength is kept fixed at $D_{2}=0.1$.
Figure 7 shows the plots of the relative weights of Subpopulation
2 and the variances of the associated probability distributions as
a function of the multiplicative noise strength $D_{1}$ for both
the Type 1 and Type 2 noises. The value of $\alpha=5.96$ the deterministic
potential of which is shown in Fig. 4. The other parameter values
are kept the same as in the case of Fig. 3. In the case of the Type
1 noise, the relative weight of a subpopulation is very little affected
by increased noise strength whereas the Type 2 noise has a substantially
greater effect in bringing about phenotypic transitions between the
stable expression states. For example, in the case of the Type 1 noise,
the change in the relative weight is very small, from 18.1\% to 18.4\%
, when $D_{1}$ is increased from 0.05 to 1.0. In the same range of
values for $D_{1}$, the change in the relative weight in the case
of the Type 2 noise is much larger, from 22\% to 47\%. In the first
case, the change in variance is also negligible, from $\backsim0.45$
to $0.51$. In the latter case, the change in variance is more prominent,
from $\backsim0.45$ to $0.95$. Similar conclusions hold true for
the other values of $\alpha$ considered in Fig. 4, namely, $\alpha=5.8$
and $\alpha=6.1$. Comparing Figs. 4 and 7, one finds that the additive
noise has the more dominant effect in the spreading of the probability
distributions. The emergent bistability model studied by us has the
major feature that the protein decay rate is nonlinear in form. The
significance of the results obtained by us is better understood if
comparisons are made with the results obtained in the cases of unregulated
gene expression and gene expression involving positive feedback (Hill
coefficient $n>1$) and linear protein dilution rate. The dynamics
of protein concentration $x$ in the case of unregulated gene expression
is given by

\begin{equation}
\frac{dx}{dt}=k_{x}-\mu x\end{equation}

where $k_{x}$ and $\mu$ denote the protein synthesis rate and the
cell growth rate (rate of increase in cell volume) respectively. Experimental
and theoretical results on the simple dynamical model indicate that
multiplicative noise in the cell growth rate accounts for a considerable
fraction of the total noise whereas the noise associated with the
synthesis rate has only a moderate contribution \cite{key-38}. In
the case of cooperative regulation of gene expression involving a
positive feedback loop but linear protein decay rate the dynamics
of protein concentration is given by

\begin{equation}
\frac{dx}{dt}=\frac{\delta+\alpha x^{n}}{1+x^{n}}-(\phi+1)x\end{equation}

If the protein synthesis term is sufficiently nonlinear ($n>1$),
one can obtain bistability in specific parameter regions. The positive
feedback amplifies the small fluctuations associated with the synthesis
rate constant so that the contribution of the synthesis term to the
total noise is not negligible compared to that of the protein decay
term \cite{key-29,key-30}. In the case of emergent bistability, we
have shown that for moderately large values of the metabolic cost
parameter $\gamma$, the multiplicative noise associated with the
maximum dilution rate parameter $\phi$ has little influence on the
shape of the steady state probability distribution. In this case,
the contribution of the protein synthesis term to the total noise
is greater. The specific nonlinear form of the protein dilution rate
appears to attenuate the effect of fluctuations in the maximum dilution
rate parameter. For low values of the parameter $\gamma$, the effect
of the multiplicative noise associated with the maximum dilution rate
on the steady state probability distribution is noticeable but the
extent of the region of bistability decreases as the $\gamma$ values
are lowered. In our study, the finding that the additive noise has the most dominant 
effect on the steady state protein distributions followed by the multiplicative 
Type 2 and Type 1 noise terms respectively is straightforward to explain. While the additive noise has a bare form, the noise in the 
effective synthesis rate constant $\alpha$ (Type 2 noise) is damped by a factor $\frac{x}{1+x}<1$ 
and the noise in the maximum dilution rate $\phi$ (Type 1 noise) is damped even further by the 
factor $\frac{x}{1+\gamma x}$ for $\gamma>1$.

\begin{figure}
\begin{centering}
\includegraphics[bb=120bp 191bp 492bp 593bp,scale=0.7]{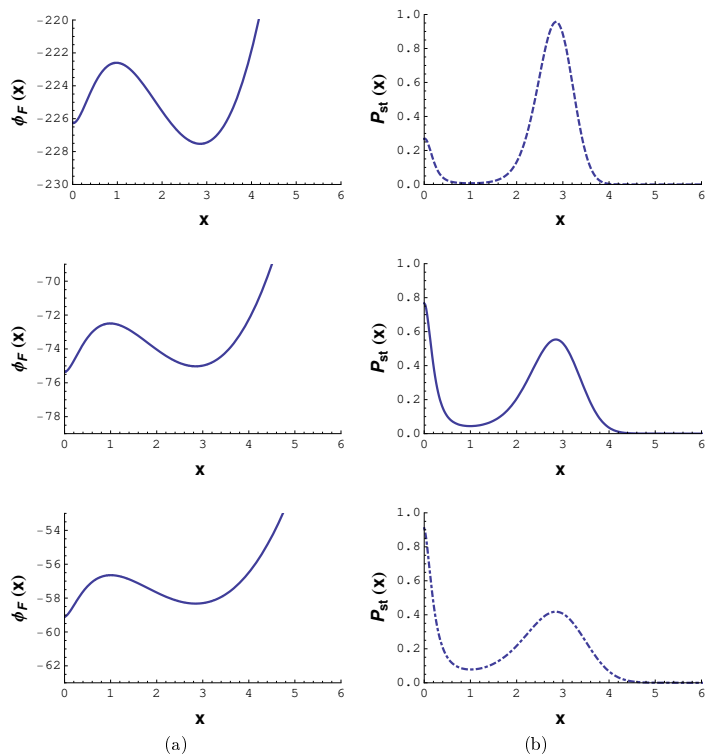}
\par\end{centering}

\centering{}\caption{Both additive and multiplicative noise terms are present with the
latter being associated with the maximum dilution rate parameter $\phi$.
The two types of noise are now correlated with $\lambda$ being the
strength of the correlation. Stochastic potential $\phi_{F}(x)$ (a)
and the steady state probability distribution $P_{st}(x)$ (b) are
displayed for $\alpha=15.34$ where the system is bistable. The other
parameter values are $\gamma=2$, $\delta=0.01$, and $\phi=20$.
The noise strengths are kept fixed at the values $D_{1}=D_{2}=0.1$
and $\lambda$ has the values $\lambda=0.7$ (dashed line), 0.0 (solid
line) and -0.7 (dot-dashed line) in the successive plots. }

\end{figure}

We lastly consider the case when the additive and multiplicative types
of noise are correlated, i.e., $\lambda$ in Eq. (3) is $\neq0$.
Such correlations occur when the two types of noise have a common
origin. We consider the multiplicative noise to be associated with
the maximum dilution rate, i.e., $g_{1}(x)$ in Eq. (2) has the form
shown in Eq. (6). Figure 8 shows the plots of the stochastic potential
$\phi_{F}(x)$ (Eq. (14)) and the SSPD $P_{st}(x)$ (Eq (12)) as functions
of $x$. The noise strengths $D_{1}$ and $D_{2}$ are kept fixed
at the values $D_{1}=D_{2}=0.1$. The stochastic potential and $P_{st}(x)$
are shown for three values of $\lambda$, $\lambda=0.7$ (dashed line),
$\lambda=0$ (solid line) and $\lambda=-0.7$ (dot-dashed line). The
$\lambda=0$ value is included for the sake of comparison between
the correlated and uncorrelated cases. The parameter values are kept
fixed at $\alpha=15.34$, $\gamma=2$, $\delta=0.01$ and $\phi=20$.
We note from Fig. 8 that negative (positive) correlation decreases
(increases) the depth of the right potential well. This is reflected
in the steady state probability distribution with negative (positive)
$\lambda$ decreasing (increasing) the height of the second peak of
$P_{st}(x)$ from that in the $\gamma=0$ case. The changes in the
depth of the left potential well and the height of the first peak
of $P_{st}(x)$ are just the reverse since the probability distribution
is normalized.

\section*{IV. MEAN FIRST PASSAGE TIMES}

We consider a bistable potential with two stable steady states at
$x_{1}$ and $x_{2}$ ($x_{1}<x_{2}$) separated by an unstable steady
state at $x_{b}$ defining the barrier state. In the presence of noise,
exits from the potential wells are possible. The exit time is a random
variable and is designated as the first passage time. In this section,
we study the effects of additive and multiplicative noise on the mean
first passage time (MFPT). Consider the state of the system to be
defined by $x$ at time $t=0$ with x lying in the interval ($a,b$).
The first passage time $T(x;\, a,b)$ is the time of first exit of
the interval $(a,b)$. The MFPT $T_{1}(x;\, a,b)$ is the average
time of the first exit and satisfies the equation \cite{key-28,key-36,key-37}

\begin{equation}
-1=A(x)\frac{dT_{1}(x)}{dx}+\frac{1}{2}B(x)\frac{d^{2}T_{1}(x)}{dx^{2}}\end{equation}

The MFPT $T_{1}(x_{1})$ for exit from the basin of attraction of
the stable steady state at $x_{1}$ satisfies Eq. (19) with the interval
$(a,\, b)=(0,\, x_{b})$ and boundary conditions given by

\begin{equation}
T_{1}^{\prime}(a;\, a,b)=0\;\mbox{{and}}\; T_{1}(b;\, a,b)=0\end{equation}

The prime denotes differentiation with respect to $x$, with reflecting
boundary condition at $a$ and absorbing boundary condition at $b$
\cite{key-28,key-36}. In a similar manner, one can compute the MFPT
$T_{1}(x_{2})$ (from Eq. (19)) for exit from the basin of attraction
of the stable state at $x_{2}$. The interval is now $(a,\, b)=(x_{b,\,}\infty)$
with $x_{b}$ and $\infty$ being an absorbing and a reflecting boundary
point respectively, i.e.,

\begin{equation}
T_{1}(a;\, a,b)=0\;\mbox{{and}}\; T_{1}^{\prime}(b;\, a,b)=0\end{equation}

\begin{figure}
\begin{centering}
\includegraphics[bb=119bp 186bp 493bp 605bp,clip,scale=0.7]{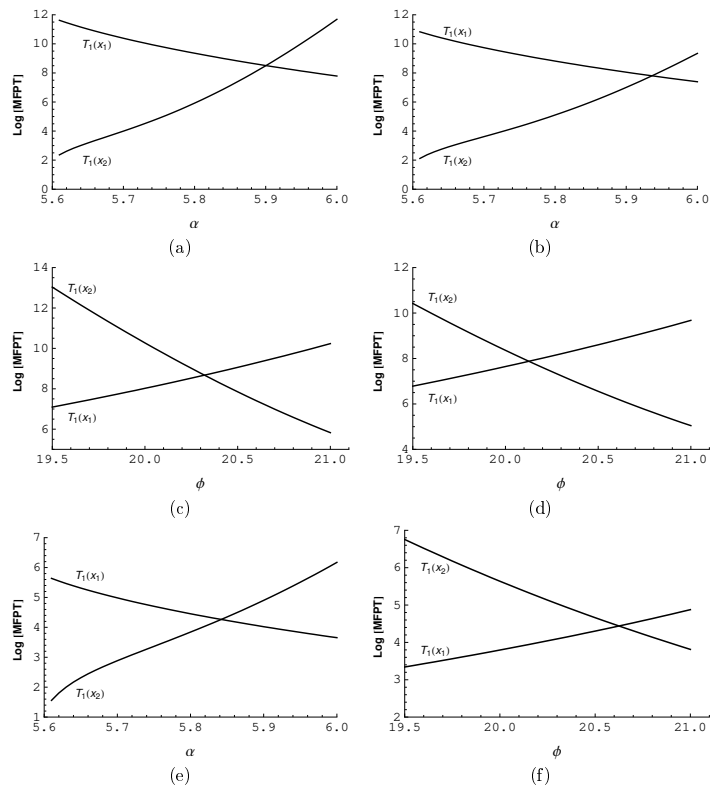}
\par\end{centering}

\centering{}\caption{Plots of logarithm of MFPT versus $\alpha$ and $\phi$. $T_{1}(x_{1})$
and $T_{1}(x_{2})$ are respectively the MFPTs for exits from the
basins of attraction of the stable steady states at $x_{1}$ and $x_{2}$.
Six different cases are considered: (a) log {[}MFPT{]} versus $\alpha$,
with only the additive noise term (strength $D_{2}=0.05$) present,
(b) log {[}MFPT{]} versus $\alpha$ with both additive and multiplicative
(associated with the effective protein synthesis rate constant $\alpha$)
noise terms are present. The noise strengths are $D_{2}=0.05$ and
$D_{1}=0.05$; (c) and (d) are similar respectively to (a) and (b)
except that log {[}MFPT{]} is plotted as a function of\textbf{ }$\phi$;\textbf{
}(e) and (f) are similar respectively to (b) and (d) except that the
additive noise strength is changed to $D_{2}=0.1$. In the cases (a),
(b) and (e), $\phi=20$. In the cases (c), (d) and (f), $\alpha=5.96$.
In all the cases the parameter $\delta=0.01$ and $\gamma=10$. }

\end{figure}

Figure 9 displays the results of the computations of the MFPTs in
different cases: (a) the logarithm of the MFPT is plotted versus $\alpha$
with only the additive noise (strength $D_{2}=0.05$) present, (b)
the same as in (a) but with the addition of the multiplicative noise
associated with the effective protein synthesis rate constant $\alpha$
(the multiplicative noise strength $D_{1}=0.05$), (c) and (d) correspond
respectively to the cases considered in (a) and (b) but now $\log${[}MFPT{]}
is plotted as a function of the parameter $\phi$ (maximum dilution
rate), cases (e) and (f) correspond respectively to (b) and (d) but
with the additive noise strength changed from $D_{2}=0.05$ to $D_{2}=0.1$.
In all the cases the parameter $\delta=0.01$ and $\gamma=10$. In
the cases (a), (b) and (e), $\phi=20$. In the cases (c), (d) and
(f), $\alpha=5.96$. Some features are worth pointing out in the plots
of the MFPTs versus $\alpha$ and $\phi$. The MFPT $T_{1}(x_{1})$
decreases and $T_{1}(x_{2})$ increases with increasing values of
$\alpha$ (Figs. 9(a), 9(b) and 9(e)) whereas $T_{1}(x_{1})$ increases
and $T_{1}(x_{2})$ decreases with increasing values of $\phi$ (Figs.
9(c), 9(d) and 9(f)). 

In the log {[}MFPT{]} versus $\alpha$ plots the rise in $T_{1}(x_{2})$
is sharper than that of $T_{1}(x_{1})$ in the log {[}MFPT{]} versus
$\phi$ plots. Similarly, in the first set of plots, the fall in $T_{1}(x_{1})$
is slower than that of $T_{1}(x_{2})$ in the second set of plots.
Comparing the Figs. 9(b) and 9(e), the crossing point of the two MFPTs
$T_{1}(x_{1})$ and $T_{1}(x_{2})$ shifts to the left when the noise
strength is changed from $D_{2}=0.05$ (Fig. 9(b)) to $D_{2}=0.1$
(Fig. 9(e)). On the other hand, the crossing point shifts to the right
in the log {[}MFPT{]} versus $\phi$ plots when the noise strength
is increased (Figs. 9(d) and 9(f)). 

A feature to emerge out of our study concerns the opposite effect
that the two types of multiplicative noise have on the dynamics. This
is more clearly seen in the plots of log {[}MFPT{]} versus $\alpha$
and $\phi$ in Fig. 9. The MFPT $T_{1}(x_{1})$ decreases and $T_{1}(x_{2})$
increases with increasing values of $\alpha$ whereas the opposite
trend is observed for increasing values of $\phi$. The shifts in
the crossing points of the two MFPTs $T_{1}(x_{1})$ and $T_{1}(x_{2})$
when the additive noise strength is increased are in opposite directions
in the two cases. Fig. 8 shows the effect of changing the correlation
strength $\lambda$ when the additive and multiplicative noise terms
are correlated. The multiplicative noise appears in the maximum dilution
rate parameter $\phi$. Similar plots (not shown) are obtained when
the multiplicative noise is associated with the effective protein
synthesis rate constant. For this case as well as for the type of
dynamics described by Eq. (18) ( with the multiplicative noise associated
with the protein synthesis rate constant), the effect of changing
the strength of the correlation between the additive and multiplicative
noises is opposite to that seen in Fig. 8. 

\begin{figure}
\centering{}\includegraphics[scale=0.7]{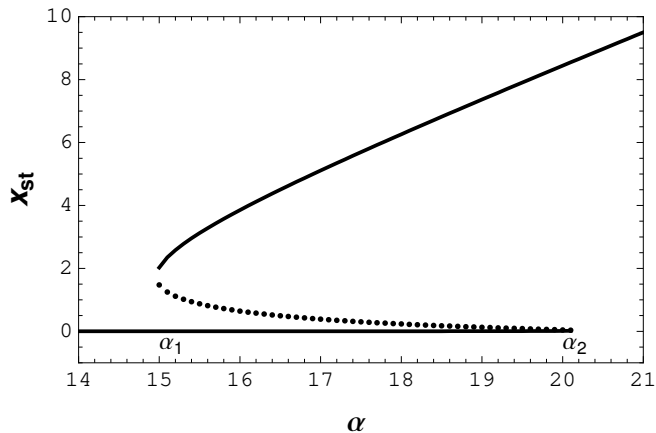}\caption{The steady state protein concentration $x_{st}$ as a function of
the parameter $\alpha$ exhibits hysteresis. The stable steady states
are represented by solid lines whereas the dotted line describes the
branch of unstable steady states. The bistable region separates the
monostable low and high expression states. The points marked $\alpha_{1}$
and $\alpha_{2}$ denote the lower and upper bifurcation points.}

\end{figure}

Bistability is often accompanied by hysteresis \cite{key-2,key-4,key-14}
an example of which is shown in Fig. 10. In the figure, the steady
state protein concentration $x_{st}$ is plotted as a function of
the parameter $\alpha$. The solid branches represent the stable steady
states separated by a branch of unstable steady states (dotted line).
The bistable region separates the monostable low from the monostable
high expression state. The transitions from the low to high and high
to low expression states are discontinuous in nature and the special
values of $\alpha$ (marked $\alpha_{1}$ and $\alpha_{2}$ in the
figure) at which they occur are the bifurcation points of the dynamics.
The path from the low to high expression state is not reversible.
As $\alpha$ increases from low to high values, a discontinuous transition
occurs at the upper bifurcation point $\alpha_{2}$. As $\alpha$
increases beyond this point, the steady state continues to be the
high expression state. If one now reverses the direction of change
in the value of $\alpha$, i.e., decreases $\alpha$ from high to
low values, there is no discontinuous transition at $\alpha_{2}$
from the high to the low expression state. The reverse transition
occurs only at the lower bifurcation point $\alpha_{1}$. The irreversibility
of paths between the low and high expression states results in hysteresis.
As pointed out earlier, in the case of the model under study,
one can pass continuously from the low to the high expression state
bypassing the region of bistability (Fig. 2(a)). An example of this
type of behavior is obtained in the study on multistability in the
lactose utilization network \cite{key-8}. In the wild-type \emph{lac}
system, one cannot go from one region of monostability to the other
without passing through a region of bistability. Appropriate modification
of the natural system make it possible to connect the two regions
of monostability via a path in which no discontinuities in the steady
state expression levels occur. Fig. 11 exhibits the steady state probability
distributions $P_{st}(x)$ versus $x$ for different values of $\alpha$
when the path from the low to the high expression state is continuous
(a) and when the transition path passes through a region of bistability
(b). In the latter case, the probability distribution is bimodal in
the intermediate range of $\alpha$ values. Only additive noise (strength
$D_{2}=0.4$) is considered in both the cases and the other parameter
values are $\delta=0.01$ and $\gamma=2$ with $\phi=1$ in (a) and
$\phi=20$ in (b). 

\begin{figure}
\begin{centering}
\includegraphics[bb=160bp 129bp 452bp 662bp,clip]{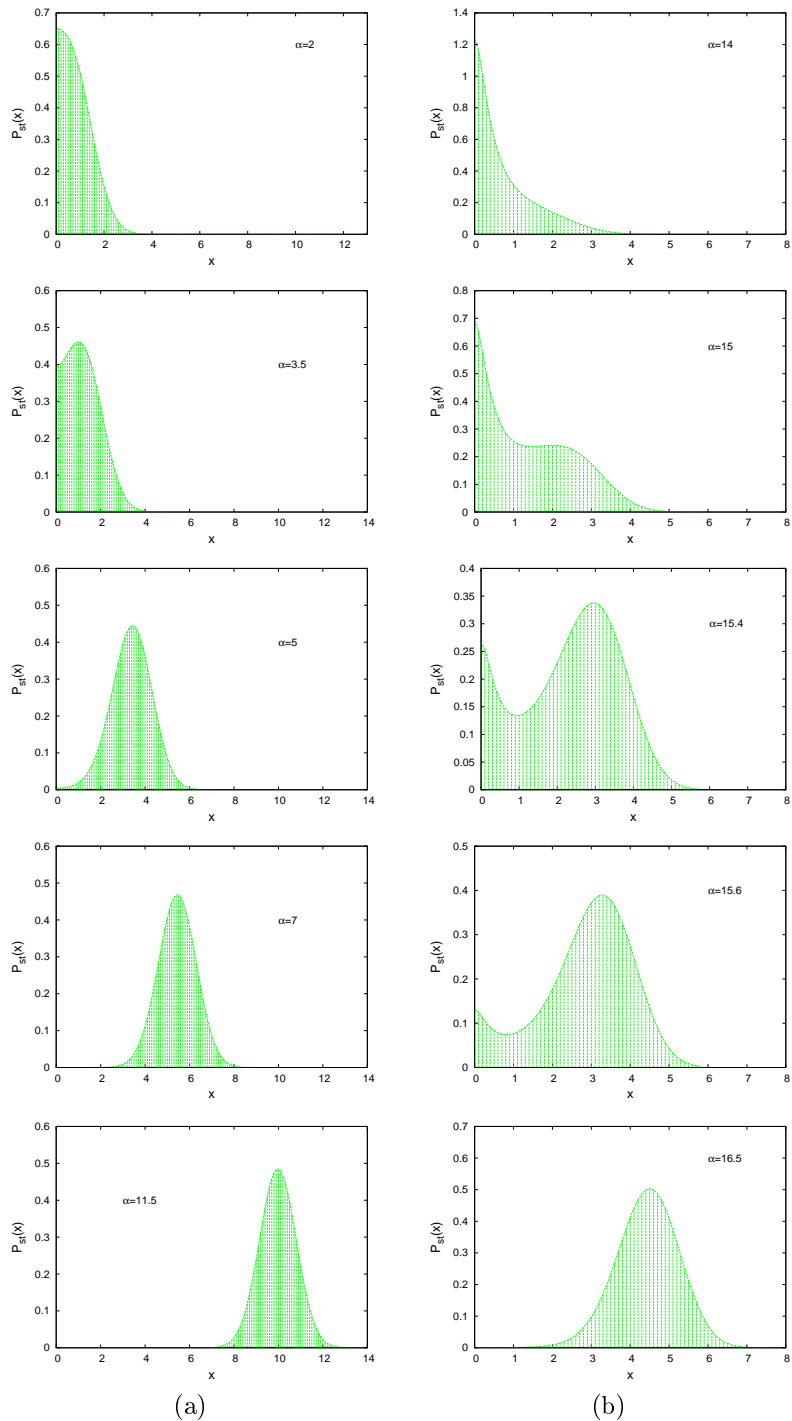}
\par\end{centering}

\caption{Steady state probability distribution $P_{st}(x)$ versus $x$. By
changing the values of $\alpha$, a transition from the low to the
high expression state is obtained in a continuous manner (a) and by
passing through a region of bistability (b). Only additive noise (strength
$D_{2}=0.4$) is considered in both the cases and the other parameter
values are $\delta=0.01$ and $\gamma=2$ with $\phi=1$ in (a) and
$\phi=20$ in (b). }

\end{figure}

\section*{V. CONCLUSION}

Emergent bistability is a recently discovered phenomenon \cite{key-16}
demonstrated in the case of a synthetic circuit. There is now some
experimental evidence that a similar mechanism may be at work in microorganisms
like mycobacteria subjected to nutrient depletion as a source of stress
\cite{key-12}. In emergent bistability, the coexistence of two stable
gene expression states is an outcome of a nonlinear protein decay
rate combined with a positive feedback in which cooperativity in the
regulation of gene expression is not essential. The nonlinear protein
decay rate is obtained if the synthesized proteins inhibit cell growth.
Cell growth results in the dilution of protein concentration so that
the protein decay rate is a sum of two terms: the dilution rate and
the protein degradation rate. In most cases, the latter rate is sufficiently
slow so that the protein decay rate is dominated by the dilution rate.
In the case of emergent bistability, the dilution rate has the form
$-\frac{\phi x}{1+\gamma x}$ where $\gamma$ is the parameter representing
the metabolic burden. For $\gamma=0$, the dilution rate is linear
in $x$ as is the protein degradation rate. In this case, bistable
gene expression via a positive feedback is possible only if the protein
synthesis rate is sufficiently nonlinear. This is achieved when the
regulatory proteins form multimers (dimers, tetramers etc. ) so that
$x$ is replaced by $x^{n}$ (n, the Hill coefficient, is $>1$) in
Eq. (1). This is the scenario that has been mostly studied so far
whereas the issue of bistability due to a combination of non-cooperative
positive feedback and nonlinear protein degradation rate is not fully
explored. 

In the case of bistable gene expression, the generation of phenotypic
heterogeneity in a population of cells is brought about by fluctuation-driven
transitions between the stable expression states. In this paper, we
analyze for the first time the effects of additive and multiplicative
noise on the dynamics governing emergent bistability. Such studies
acquire significance in the light of the fact that the generation
of phenotypic heterogeneity enables a subset of a population of microorganisms
to survive under stress. Examples of such stresses include depletion
of nutrients, environmental fluctuations, lack of oxygen, application
of antibiotic drugs etc. There is now considerable experimental evidence
that positive feedback and gene expression noise provide the basis
for the `advantageous' heterogeneity observed in microbial populations
\cite{key-10,key-24,key-25}. The heterogeneity is usually in the
form of two distinct subpopulations with low and high expression levels
of a key regulatory protein, e.g., ComK in \emph{B. Subtilis} \cite{key-10,key-25}
and Rel in \emph{M. Smegmatis }\cite{key-11,key-12}. High ComK levels
in a fraction of the \emph{B. Subtilis }population result in the development
of `competence' in the subset of cells enabling the subpopulation
to adapt to changed circumstances. The role of noise in bringing about
phenotypic transitions from the low to high ComK expression states
has been demonstrated experimentally \cite{key-25}, the reduction
of noise results in a smaller fraction of cells in which competence
is developed. In mycobateria, high Rel levels in a subpopulation of
cells (the so-called \emph{persisters}) initiate the stringent response
in these cells enabling the subpopulation to survive under stresses
like nutrient depletion. The role of positive feedback and gene expression
noise in the generation of two distinct subpopulations has been investigated
experimentally in \emph{M. Smegmatis }\cite{key-11,key-12}. As mentioned
earlier, there is some experimental evidence \cite{key-12} that the
appearance of two distinct subpopulations, in terms of the Rel expression
levels, is an outcome of emergent bistability. The key elements of
the stringent response pathway and the ability to survive over long
periods of time under stress are shared between the mycobacterial
species \emph{M. Smegmatis }and \emph{M. Tuberculosis }\cite{key-41,key-42,key-43}.
The latter, the causative agent of tuberculosis, has remarkable resilience
against various types of stress including that induced by drugs. A
mechanism similar to that in \emph{M. Smegmatis }may be responsible
for the generation of the subpopulations of persisters (not killed
by drugs) and non-persisters. Studies based on stochastic dynamic
approaches provide knowledge of the key parameters controlling the
operation of relevant gene circuits and the effects of fluctuations
in these parameters towards the generation of phenotypic heterogeneity.
The studies provide valuable inputs in the designing of effective
strategies for drug treatment.

We have further
pointed out the possibility of connecting the low and high expression
states in a continuous manner. In this case, the region of bistability
is bypassed so that no discontinuity in the steady state protein levels
as seen in the hysteresis curve of Fig. 10 occurs. The bypassing
 is facilitated for low values of the `metabolic burden' parameter $\alpha$. Since experimental
modulation of the parameter $\alpha$ and $\phi$ are possible \cite{key-16},
the theoretical prediction could be tested in an actual experiment.
The dynamics considered in the present paper correspond to that of
an average cell. In a microscopic model, the growth rates of the two
subpopulations in the region of bistability could be different \cite{key-16}.
The potential impact of the growth rate on model parameters other
than the protein dilution rate has been ignored in our study but this
simplification is possibly well justified \cite{key-18}. An issue
that has not been explored in the present study is that of stochastic
gene expression resulting in a bimodal distribution of protein levels
in the steady state but without underlying bistability arising from
deterministic dynamics. The issue of bimodality without bistability
has been explored both theoretically \cite{key-6,key-37} and experimentally
\cite{key-44}. The present study is based on the approximate formalism
involving the Langevin and FP equations. The chief advantage of the
formalism lies in its simplicity and its ability to identify the separate
effects of additive and multiplicative noises of various types. The
formalism is valid when the number of molecules involved in the dynamics
is large and the noise is small. Studies based on more rigorous approaches
are desirable for a greater understanding of the effects of noise
on emergent bistability.

\section*{ACKNOWLEDGMENT}

SG acknowledges the support by CSIR, India, under Grant No. 09/015(0361)/2009-EMR-I.

\end{document}